# Integrating Notch Filtering and Statistical Methods for Improved Cardiac Diagnostics Using MATLAB

Lohit Bibar
BTech (ECE) Student
Department of Electronics & Communication Engineering,
Dr. BC Roy Engineering College,
Durgapur, West Bengal, India
lohitbibar2020@gmail.com

Samali bose
BTech (ECE) Student
Department of Electronics & Communication Engineering,
Dr. BC Roy Engineering College,
Durgapur, West Bengal, India
bosesamali2003@gmail.com

Tribeni Prasad Banerjee
Department of Electronics & Communication Engineering,
Dr. BC Roy Engineering College,
Durgapur, West Bengal, India

***Abstract—*** *A Notch Filter is essential in ECG signal processing to eliminate* narrowband noise, *especially* powerline interference *at 50 Hz or 60 Hz.* ***This interference overlaps with vital ECG signal features, affecting the accuracy of downstream*** *classification tasks (e.g., arrhythmia detection). A properly designed notch filter enhances signal quality, preserves essential ECG components (P, QRS, T waves), and improves the performance of machine learning or deep learning models used for ECG classification.*

## 1. INTRODUCTION

The electrical activity of the heart is represented by an electrocardiogram signal. All heart diseases can be understood using these heart signals. Basically, the heart signal that is known from electrocardiograph, seeing that the experienced doctor tells what happened to the patient and what treatment should be done. More than 30years ago, this technology was not advanced significantly. It is very difficult for doctors to see the electrocardiogram signal and tell 'actual disease because any heart electrocardiogram signal is too difficult to understand. According to the world health organization (who), cardiovascular diseases (cvds) are the leading cause of death globally, accounting for an estimated 17.9 million lives each year. Cvds are a group of disorders of the heart and blood vessels, including coronary heart disease, cerebrovascular disease, rheumatic heart disease, and other conditions. More than four out of five cvd deaths are due to heart attacks and strokes, and one-third of these deaths occur prematurely in people under 70 years of age. Thus, gradually, with the advent of technology (such as ai and machine learning), understanding has improved. Many researchers have studied improved classification and filtering techniques for electrocardiograms. Electrocardiogram (ECG) signals play a crucial role in detecting cardiac abnormalities. Various techniques, such as Fourier Transform (FFT), Short-Time Fourier Transform (STFT), and statistical methods, have been employed to enhance feature extraction for ECG classification. However, ECG signals often suffer from distortions that must be carefully addressed to improve classification accuracy. Traditional ECG signal decomposition methods often rely on the Fourier Transform, which minimizes squared errors but struggles with outliers and overlapping frequency components [2]. An ECG classification approach leveraging deep transfer learning with ResNet-18 and Short-Time Fourier Transform (STFT) was introduced by Cao et al. (2023), aiming to enhance feature extraction and classification accuracy [4],[7]. Harmonic distortion is a well-known issue in industrial power systems, affecting power quality and signal integrity, as analyzed by Riaz et al. (2021) [11]. Notch filters, or band-stop filters, are commonly used in signal processing to reduce a specific range of unwanted frequencies, certainly noise signals, while allowing the rest of the signal to remain largely unaffected. Notch filters are essential in signal processing for attenuating specific frequency bands while preserving others. Initially passive, using resistors, capacitors, and inductors, their limitations led to the development of active filters incorporating operational amplifiers (op-amps). Modern designs integrate low-pass and high-pass filters with a summing amplifier to enhance selectivity and performance [13]. They are particularly effective at removing noises like the 50/60 Hz hum from power lines in audio and communication systems. For instance, let us consider an ECG signal. Conventional fixed-notch filters lack adaptability to interference changes. Adaptive notch filters dynamically adjust to input signals, effectively suppressing noise while preserving essential ECG signal features for improved performance [17]. Power line interference poses a challenge in obtaining clear ECG signals. Analog notch filters help attenuate unwanted frequencies, while digital FIR equiripple filters further refine the signal, ensuring minimal distortion and optimal interference suppression [16]. Traditional ECG processing used analog filters, which lacked adaptability to noise variations. Combining analog filtering with digital rectangular window techniques enhances flexibility, precision, and artifact suppression, improving the signal-to-noise ratio for clearer ECG readings [15]. Notch filters are typically created by combining low-pass and high-pass filters that function at complementary frequencies. This combination allows for the attenuation of frequencies within a specified frequency range while maintaining the quality of signals outside that range. Filter circuits are crucial in signal processing, but traditional active-RC filters suffer from noise due to resistors, capacitors, and op-amps. The study highlights that all-pole active-RC filters with minimal sensitivity to component tolerances exhibit lower noise levels. Impedance tapering, a key design methodology, reduces sensitivity by scaling RC ladder sections, optimizing component values, and using a single op-amp per circuit, ensuring low noise, reduced power consumption, and cost-effectiveness [20]. The low-pass filter (LPF) demonstrates a cutoff frequency within an acceptable tolerance range, aligning with theoretical expectations. In this setup, the low-pass filter (LPF) permits frequencies below the notch to pass, while the high-pass filter (HPF) allows frequencies above the notch to pass. By carefully designing these filters to meet at a specific centre frequency, the targeted unwanted frequency band is effectively attenuated, forming the notch in the frequency response. Op-amp-based notch filters are highly adaptable and are used in many applications. They help reduce noise from specific sources in audio processing while preserving the fidelity of music or speech. In medical instrumentation, these filters eliminate interference in critical measurements like ECG and EEG signals. Furthermore, in communication systems, they are essential for filtering out narrow-band interference, leading to clearer signal transmission. The integration of LPF and HPF using op-amps results in efficient, flexible, and stable notch filter designs that can be customized for various needs. The conventional notch filter remains a reliable and widely supported solution for suppressing power line interference in wearable healthcare devices. Its effectiveness and established design make it a preferred choice in many applications. However, by reconfiguring its requirements using the Twin-T approach, we enhance its practicality, leveraging its simplicity while addressing integration challenges. This refined design balance's traditional reliability with improved adaptability for modern circuit implementations.

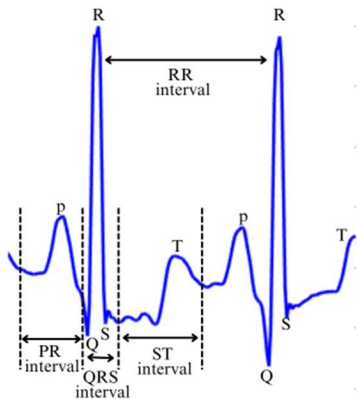

Figure 1. ECG Signal with PQRST wave

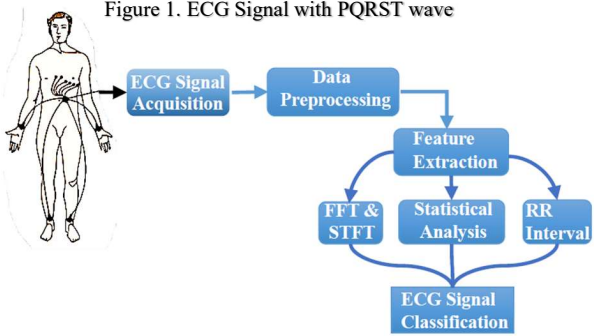

Figure 2. A block diagram of proposed method's stage.

## 2. METHODOLOGY:

This section presents a comprehensive overview of the experimental setup and ECG data collection procedures. It details the filtering techniques applied to mitigate heavy noise in ECG signals and introduces an adaptive Notch Filter for enhanced signal clarity. The study further explores ECG signal segmentation, including disease identification, and incorporates STFT analysis for feature extraction from each segment. Various classification techniques are employed to assess the effectiveness of the analysis, followed by statistical analysis to evaluate the performance and reliability of the proposed methods.

### DATA COLLECTION:

Physionet offers several ECG databases. Our research utilized ECG data from various physionet databases, including MIT-BIH svdb, Cudb, MIT-BIH nsrdb, and MIT-BIH Arrhythmia Database (mitdb). We examined these datasets to identify potential variations in ECG signal characteristics. The records in each dataset contain different sample sizes: MIT-BIH svdb has 230400 samples, cudb has 127,232 samples, MIT-BIH nsrdb has 11730944 samples, and mitdb has 650000 samples. These datasets represent diverse subject groups and recording conditions, with sampling rates ranging from 128 Hz to 360 Hz and varying levels of interference. We used ECG1 data from all records without exclusion. The onset of ventricular arrhythmia can be detected by analyzing Heartbeat rate (HBR), R-R intervals, and QRS amplitudes.

### DATA PREPROCESSING:

We successfully implemented a dual-filtering technique to enhance ECG signal quality and enable precise feature extraction. This approach integrates a notch filter and a Butterworth band-pass filter to effectively mitigate power-line interference, baseline drift, and high-frequency noise, optimizing the ECG signal for analysis. First, we applied a notch filter specifically designed to attenuate 40 Hz power-line interference, which can obscure critical ECG features. This filter selectively suppressed the targeted frequency while preserving the integrity of the desired signal range. Following this, a Butterworth band-pass filter with a frequency range of 0.5 to 40 Hz was implemented. This filter served two primary purposes: eliminating low-frequency baseline wander and reducing high-frequency noise from sources such as muscle activity, motion artifacts, electromagnetic interference, poor electrode contact, and human voice disturbances. The Butterworth filter was chosen for its smooth frequency response and minimal phase distortion—key attributes for biomedical signal processing. Using MATLAB, we meticulously designed the notch filter with optimized parameters: a lower cutoff frequency of 0.5 Hz, an upper cutoff frequency of 40 Hz, and a filter order of 3. These settings ensured a maximally flat frequency response within the desired range, preserving ECG signal fidelity. This dual-filtering process significantly improved ECG signal quality by reducing noise and artifacts outside the frequency range of interest. It effectively minimized low-frequency baseline wander caused by respiration or patient movement, ensuring that only relevant cardiac information remains in the lower frequency spectrum. Simultaneously, it attenuated high-frequency muscle noise, enhancing the clarity of ECG waveforms, particularly the PQRST complex. The combined application of these filters proved to be a robust preprocessing method for ECG data. While the notch filter targeted specific frequency components, such as the 40 Hz power-line noise, the Butterworth band-pass filter ensured comprehensive signal refinement across a broader spectrum. This systematic noise reduction approach not only enhanced the visual clarity of ECG signals but also improved the reliability and accuracy of feature extraction processes, including PQRST complex detection and frequency-domain analysis.

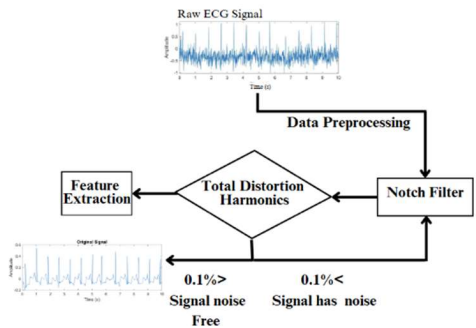

For smoother signal processing in ECG analysis, a **notch filter** is essential to remove unwanted noise, particularly **powerline interference at 50/60 Hz**. This interference can obscure critical cardiac signals, leading to inaccurate diagnostics. A notch filter selectively attenuates this narrowband noise while preserving the essential ECG components.

By integrating **filtering techniques and spectral methods** in **MATLAB**, we can enhance ECG signal quality, improve feature extraction, and ensure more reliable cardiac diagnostics. MATLAB's robust filtering toolbox allows precise notch filter design, ensuring minimal distortion of vital ECG waveforms.

## 3. DESIGNING NOTCH FILTER OF SUITABLE ORDER.

### LOW-PASS FILTERS

A low-pass filter (LPF) is an electronic circuit designed to permit signals with frequencies lower than a specified cutoff frequency to pass while reducing the amplitude of higher-frequency signals. These filters find numerous applications, including audio processing, telecommunications, and signal conditioning.

Parameters of Low-Pass Filters:
- Cutoff Frequency ($f_c$): This is the frequency at which attenuation of the input signal begins. Frequencies below this point can pass through with minimal loss, whereas frequencies above will be gradually diminished.
- Passband: This is the range of frequencies below the cutoff point where the output signal remains relatively unchanged. It is typically characterized by a defined gain, usually at 0 db.
- Types: lpfs can be built using either passive components (like resistors and capacitors) or active components (such as op-amps), with active filters generally providing superior performance in terms of gain and stability.

HOW TO DESIGN THE LOW-PASS FILTER REGION?
The low-pass filtering action is primarily achieved through the combination of R1, C2, and C3. These components create a network that allows low-frequency signals to pass while attenuating higher-frequency components. The capacitor C2 provides a path for high-frequency components to be shunted, thereby reducing their amplitude. The presence of resistor R1 ensures that low-frequency signals experience minimal resistance, allowing them to propagate through the network

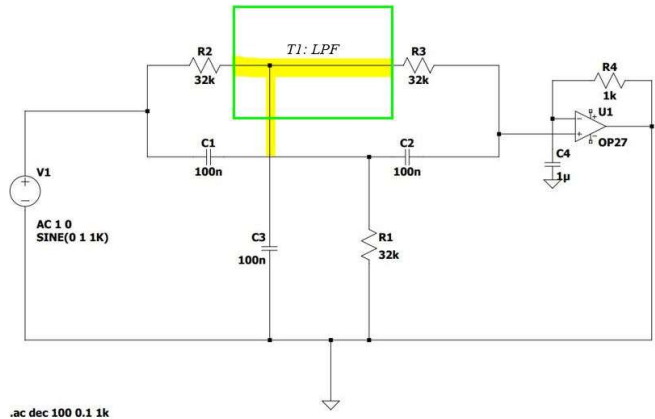

Twin-T Notch filter circuit: Low Pass Filter region circuit diagram

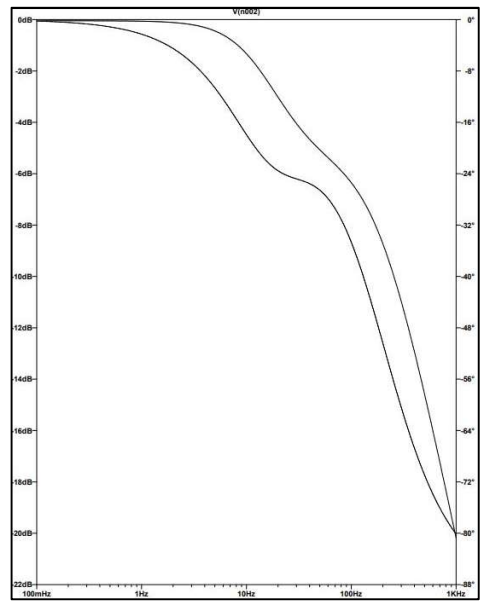

Trace analysis of LPF region in the Twin-T circuit

FILTER TYPE
- It is a low-pass filter (LPF).
- The passband (low frequency) shows 0Db gain.
- The stopband (high frequency) shows attenuation.

**HIGH-PASS FILTERS**

A high-pass filter (HPF) is an electronic circuit designed to allow signals with frequencies higher than a specified cutoff frequency to pass while reducing the amplitude of lower-frequency signals. These filters are widely used in audio processing, telecommunications, and signal conditioning.
Parameters of High-Pass Filters:
- Cutoff Frequency (fc): This is the frequency at which attenuation of the input signal begins. Frequencies above this point pass through with minimal loss, while frequencies below are gradually diminished.
- Passband: This is the range of frequencies above the cutoff point where the output signal remains relatively unchanged. It is typically characterized by a defined gain, usually at 0 db.
- Types: hpfs can be built using either passive components (like resistors and capacitors) or active components (such as op-amps), with active filters generally providing superior performance in terms of gain and stability.

HOW TO DESIGN THE HIGH-PASS FILTER REGION?
The high-pass filtering behavior is established by the combination of R2, R3, and C1. This section attenuates low-frequency signals while allowing higher-frequency signals to pass. The capacitor C1 blocks DC and low-frequency components, while resistors R2 and R3 set the frequency response of this high-pass network. The interaction of these elements ensures that signals above a certain cutoff frequency are transmitted with minimal attenuation.

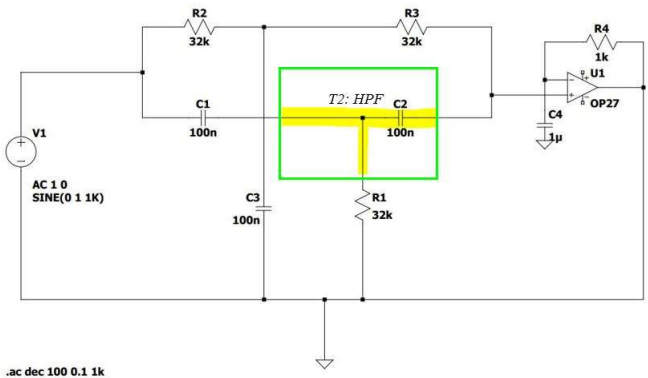

Twin-T Notch filter circuit: High Pass Filter region circuit diagram

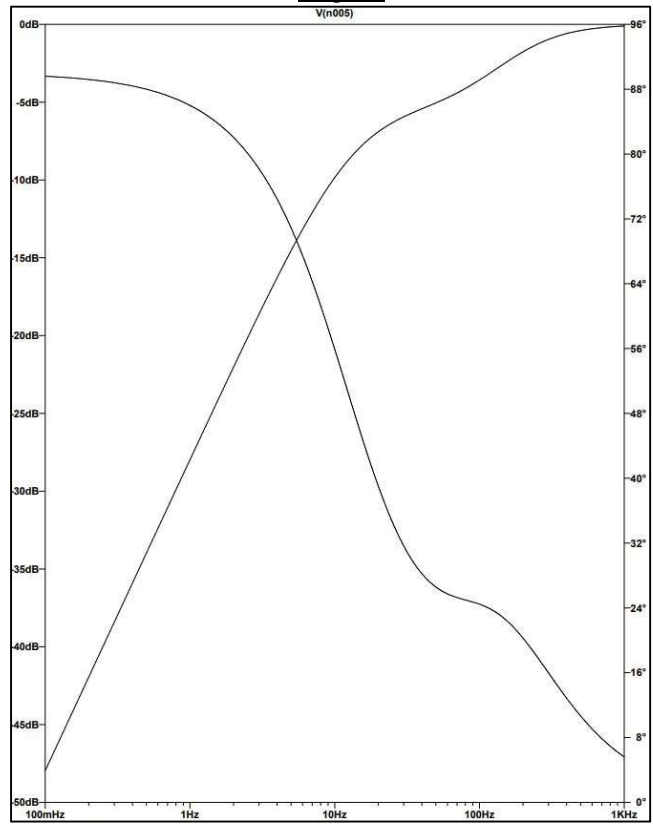

Trace analysis of HPF region in the Twin-T circuit

FILTER TYPE
- This is a high-pass filter.
- The passband (high frequency) shows a flat gain of 0 db.
- The stopband (low frequency) attenuates the signal significantly.

**INTEGRATED NOTCH CIRCUIT:**
**TWIN-T CONFIGURATION**
**I.     Integration into a Notch Filter**

The combination of the low-pass and high-pass sections results in a notch filter response, which is characterized by significant attenuation at a specific notch frequency. The twin-T configuration, comprising C1, C2, C3, R1, R2, and R3, establishes a frequency-selective network that rejects a narrow band of frequencies centered around the notch frequency. The operational amplifier U1, configured as a buffer, ensures that the filter's performance is not degraded by loading effects while maintaining a stable output. The resistor R4 provides the necessary feedback to stabilize the amplifier's operation.

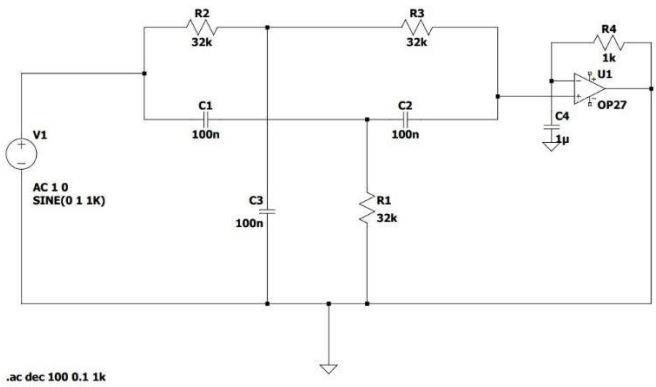

Notch Filter having Twin-T configuration within the cutoff range 45-50Hz

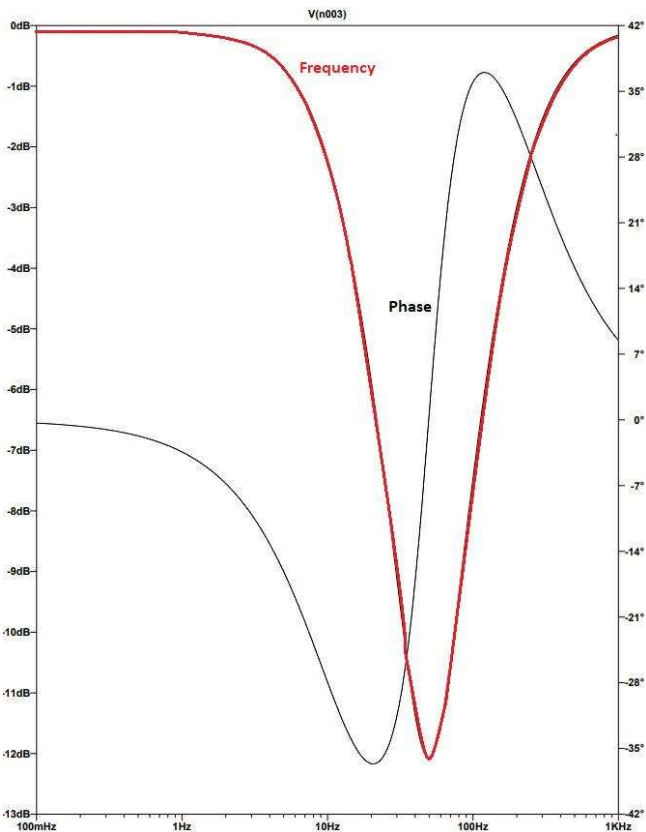

## II. Frequency Response and Performance

The circuit is designed to attenuate frequencies in the range of 40-50 Hz, which can be verified through AC analysis simulations. The chosen component values dictate the notch frequency, calculated using the standard twin-T notch filter formula:

$F_{notch}$ or $f_{cutoff} = \frac{1}{2\pi RC}$

Where R = 32kω and C = 100nf.

Substituting these values, the theoretical notch frequency is determined as:
$F_{notch}$ or $f_{cutoff}$ = 49.7 Hz

This aligns well with the expected rejection range of 40-50 Hz, making the circuit suitable for attenuating power line interference or other unwanted signals in this frequency band.

## III. Order of the filter

Theoretically, a second-order filter has a slope of -40 db/decade per pole. The twin-T notch filter consists of two RC networks, each contributing a first-order response. Since the twin-T configuration includes two capacitors and two resistors in the signal path, it results in a second-order response.

Thus, the circuit exhibits second-order behavior, leading to sharper attenuation around the notch frequency compared to a first-order filter.

## IV. Transfer Function of the Twin-T Notch Filter

The given circuit is a **Twin-T notch filter**, and its transfer function can be derived based on the component values and circuit topology.

Step 1: Define the Twin-T Network Components
The passive network consists of the following elements:
- Resistors**:** R1=R2=R3=R
- Capacitors**:** C1=C2=C3=C

Step 2: General Transfer Function of a Twin-T Notch Filter
The standard transfer function for a Twin-T network is:

$$H(s) = \frac{S^2 + \left(\frac{1}{RC}\right)S}{S^2 + \left(\frac{1}{RC}\right)S + \left(\frac{1}{R^2C^2}\right)}$$

Where s is the complex frequency variable, s=jω
At the notch frequency $f_{notch}$, the denominator term ensures complete attenuation at:

$F_{notch}$ or $f_{cutoff} = \frac{1}{2\pi RC}$

Step 3: Substituting Given Values
Given that R=32kω =32×10³Ω and C=100nf =100×10⁻⁹, we calculate:

$F_{notch}$ or $f_{cutoff} = \frac{1}{2\pi \times R \times C}$
$= \frac{1}{2\pi \times 32 \times 10^3 \times 100 \times 10^{-9}}$
$\approx 49.7 \text{Hz}$

Thus, the transfer function simplifies to:

$H(s)$
$= \frac{S^2 + \left(\frac{1}{32 \times 10^3 \times 100 \times 10^{-9}}\right)S}{S^2 + \left(\frac{1}{32 \times 10^3 \times 100 \times 10^{-9}}\right)S + \left(\frac{1}{32 \times 10^{3^2} 100 \times 10^{-9^2}}\right)}$

$$H(s) = \frac{S^2 + (312.5)S}{S^2 + (312.5)S + (9.77 \times 10^4)}$$

Where, s=jω

This transfer function confirms that the circuit attenuates signals at $f_{notch} \approx 49.7$Hz, making it a **second-order notch filter**.

In conclusion, we see that through the combination of a low-pass filter, a high-pass filter, and an active buffering stage, the given circuit effectively implements a notch filter. The

Twin-T configuration ensures selective frequency attenuation, while the operational amplifier enhances circuit stability and performance. This design is particularly useful for applications requiring the suppression of narrowband interference within the targeted frequency range.

| Table: Parameters of Twin-T Notch filter | | | |
|---|---|---|---|
| **Type of T-Section** | **Resistor (R)** | **Capacitor (C)** | **Feedback Resistors** |
| LPF (Low Pass Filter) region | 2 resistance components (32kω, 32kω) | 1 capacitance component (100nf) | NA |
| HPF (High-Pass Filter) region | 1 resistance component (318kω) | 2 capacitance components (100nf,100nf) | NA |
| Amplifier region (consists a ±12V Op-Amp) | NA | 1μf | 1kω |

Component table

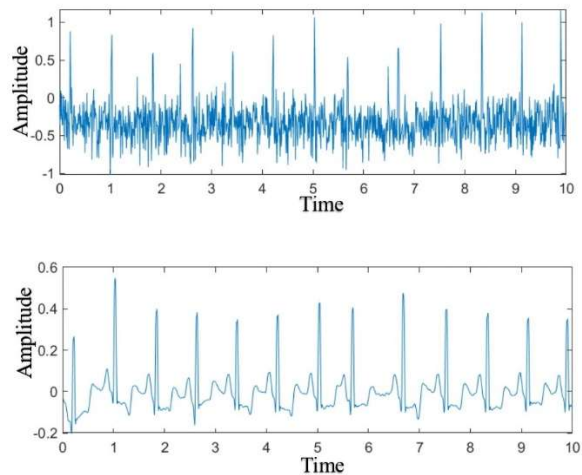

Figure 8. Raw ECG Noisy Signal VS Filter ECG Signal

**Total Harmonic Distortion (THD) :**

Total Harmonic Distortion (THD) is a crucial metric for assessing ECG signal quality after filtering, ensuring the integrity of processed data. The accuracy of ECG signal classification heavily depends on signal clarity, with THD serving as a key indicator of data quality. The Powergui block enables real-time THD analysis, allowing dynamic evaluation of signal fidelity following filtration. THD quantifies signal distortion by comparing the total power of harmonic frequencies to that of the fundamental frequency.

Calculate THD:
- Compute the power of the fundamental frequency and the power of the harmonics.
- Use the formula for THD:

$$\text{THD} = \frac{\sum_{n=2}^{\infty} V_n^2}{V_1} * 100\%$$

Where $V_1$ Is the RMS value of the fundamental frequency, and $V_n$ Is the RMS value of the n-th harmonic.

**ECG Signal Representation**

A spectrogram visualizes the intensity of a signal across different frequencies, highlighting those with the highest energy and illustrating how this energy changes over time. A two-dimensional histogram provides a visual representation of the distribution of pixel values within an electrocardiogram (ECG) signal (Figure). The Short-Time Fourier Transform (STFT) is used to analyze the ECG signal over time, offering a time-frequency representation. These analyses were performed using MATLAB. This technique is particularly useful for non-stationary signals like ECG, where the frequency content can change unexpectedly due to various physiological events, such as arrhythmias. The STFT applies a sliding window to the signal, enabling an examination of how frequency components evolve. The STFT was plotted using (1).

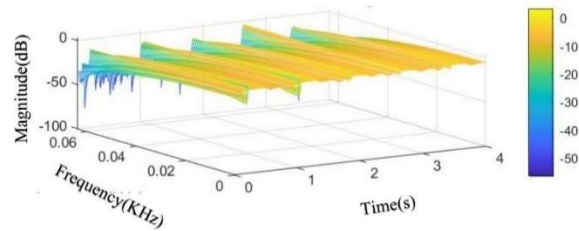

Figure 10. STFT 3D PLOT of ECG Signal

We used a
Hamming window of length 4, calculated by the following formula:

| Number | Window | Formula |
|--------|--------|---------|
| 1 | Hamming | $W[n] = 0.54 - 0.46\cos(\frac{2\pi n}{4-1})$ .... (1) |
| 2 | Kaiser | $W[n] = \frac{I_0\left(\beta\sqrt{1-(\frac{2n}{N}-1)^2}\right)}{I_0(\beta)}$ |
| 3 | Blackman | $W[n] = 0.42 - 0.5\cos\left(\frac{2n\pi}{N}\right) + 0.08\cos\left(\frac{4n\pi}{N}\right)$ |
| 4 | Gaussian | $W[n] = e^{-0.5\left(\frac{n-\frac{N}{2}}{\sigma\frac{N}{2}}\right)^2}$ |

Table 3. All Window Formula Of STFT

### Feature Extraction

The statistical analysis and frequency-based features of ECG signals are extracted for further evaluation. Detecting the R-R interval is a critical step in ECG analysis, as it provides vital insights into heart rate and rhythm. The R-R interval, defined as the time between consecutive R-wave peaks in the ECG, serves as a key indicator of cardiac health and variability. This section outlines the methodology used for precise R-R interval detection and subsequent feature extraction. The Pan-Tompkins algorithm is employed to process ECG signals, identifying QRS complexes with a focus on R-wave peaks, which are crucial for R-R interval calculation."

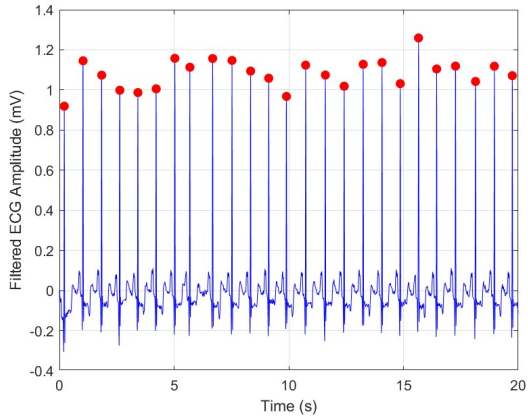

Figure 11. R Detection of ECG Signal

| Number | Feature | Formula |
|--------|---------|---------|
| 1 | Mean | $\mu = \frac{1}{N}\sum_{i=1}^{N} X_i$ |
| 2 | Standard deviation | $\sigma = \frac{1}{N}\sqrt{\sum_{i=1}^{N}(X_i - \mu)^2}$ |

Table 4. Statistical Features Derived Formula

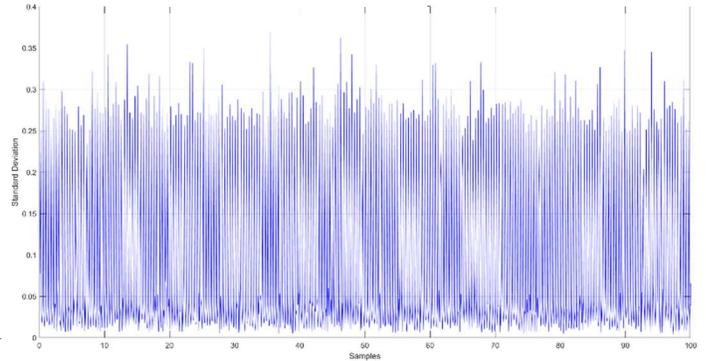

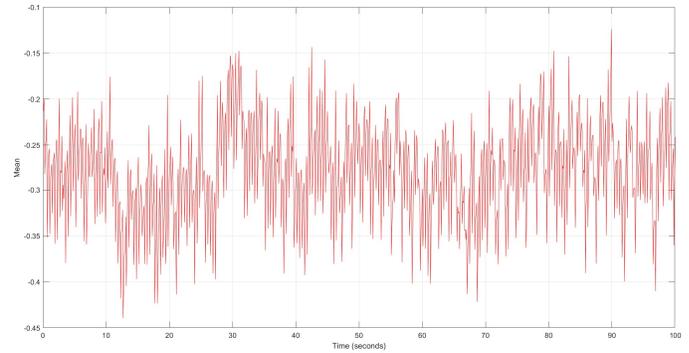

Figure 12. Standard Deviation of ECG Signal

Next, we perform a Fast Fourier Transform (FFT) for each windowed segment. For simplicity, we perform a 2-point FFT for each segment. For a N-point signal $x = [x_0, x_1, x_2, x_3...]$

The N-point FFT is given by:

$X[K] = \sum_{n=0}^{N} x[n] \cdot e^{-j\frac{2\pi Kn}{4}}$
For k = 0,1,2,3
We compute the FFT for each windowed segment

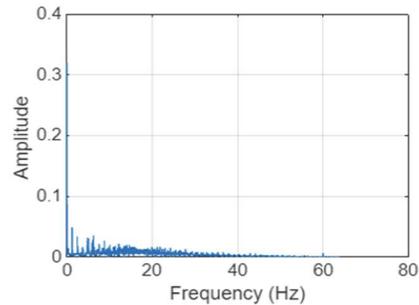

Figure 13. FFT of ECG Signal

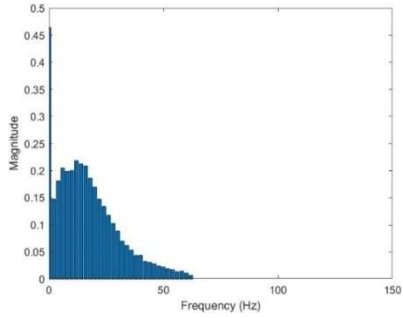

Figure 9. Histogram of ECG Signal

### ECG Signal Classification

As delineated in the introductory section, ECG signals are categorized into five distinct classes: one normal (N) and four abnormal (VEB, SVEB, FB, Q). Subsequent to the preprocessing stage and feature extraction from individual signals, classification is conducted accordingly.

### Normal Signal

| Segment | Represent |
|---------|-----------|
| PQ | Atrial depolarization |
| QRS | Ventricular depolarization |
| ST | Ventricular repolarization |

Table 5.

Normal Heart Beat 60-90 bpm.

| Features | Duration |
|----------|----------|
| PR | 0.12 - 0.20 sec |
|  |  |
| QRS | 0.06 - 0.10 sec |
| ST | ST ≤ 0.40 sec |

Table 6. Normal Duration of ECG Signal

### Unknow Signal

Our segmentation technique accurately detects the P, Q, R, S, and T waves. By focusing on the RR interval with the Pan-Tomkins Algorithm, QRS duration, and QT interval. Then, STFT is applied to every segment for better time-frequency analysis, helping to observe transient frequency changes in which we extract features essential for arrhythmia diagnosis. All segments are plotted in Different window Function. Several window functions are commonly used in STFT, each with unique characteristics, such as the short-time Fourier transform (STFT) evaluation of ECG indicators, and the selection of window function performs a critical function in balancing time and frequency resolution. The square (boxcar) window is the best, presenting the first-rate time resolution but bad frequency decision due to big spectral leakage, making it less suitable for a particular frequency analysis. The Hamming window is normally used for ECG alerts because it reduces spectral leakage with smooth tapering, offering a very good compromise between time and frequency decisions. Similarly, the Hanning (Hann) window offers a moderate spectral leakage discount and is frequently used for trendy-frequency analysis in ECG alerts. The Gaussian window provides an incredible frequency decision, formed like a bell curve, making it perfect for cases in which frequency precision is prioritized over time resolution. Lastly, the Blackman window gives even lower spectral leakage than the Hamming and Hanning windows, taking into account the very precise frequency separation, which is particularly beneficial in studying subtle adjustments in heart rate variability. Each window function has precise traits, and the choice depends on the specific necessities of the ECG analysis, which includes the want for detecting fast temporary events or analyzing longer-time period frequency additives

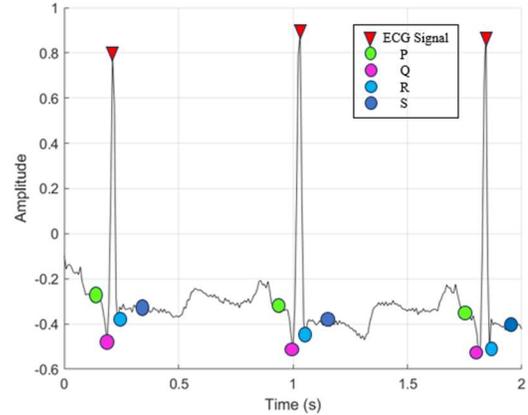

Figure 14. Detect PQRST of ECG Signal

### 3. RESULT AND DISCUSSION

This section presents the findings of our experimental study. To enhance analysis and classification, we utilized the Short-Time Fourier Transform (STFT) for accurate frequency visualization of electrocardiogram (ECG) signals and Total Harmonic Distortion (THD) to assess ECG signal quality. Additionally, this research explores the effectiveness of various STFT window functions, including Hamming, Kaiser, Blackman, and Gaussian, for improved frequency representation of ECG signals. The results demonstrate that STFT provides valuable insights, highlighting its potential in ECG signal classification. Furthermore, this study investigates the classification of ECG signals into normal and abnormal categories using Fast Fourier Transform (FFT) for feature extraction. FFT is employed to derive relevant frequency-domain attributes, thereby improving classification accuracy. The findings suggest that FFT-based feature extraction significantly contributes to the reliable identification of cardiac conditions. In addition to frequency-domain analysis, statistical analysis was performed on ECG signals to extract key features such as mean, variance, standard deviation, . These statistical parameters provide insights into signal distribution, variability, and underlying patterns, further enhancing the classification process. The integration of statistical features with frequency-domain attributes improves the robustness and reliability of ECG signal classification.

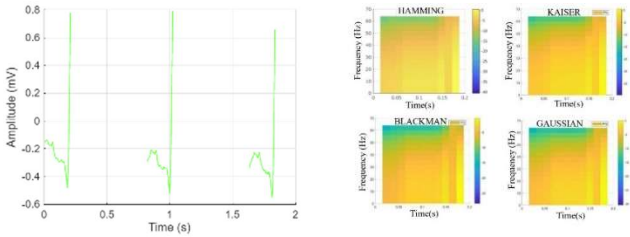

Figure 15. PQ Segment of ECG Signal

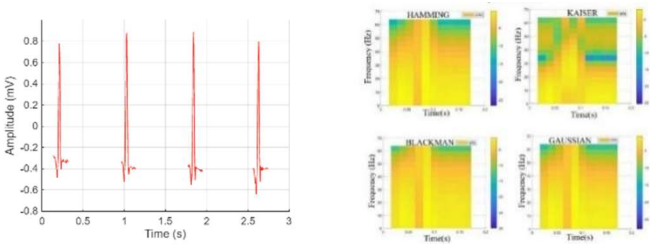

Figure 16. QRS Segment of ECG Signal

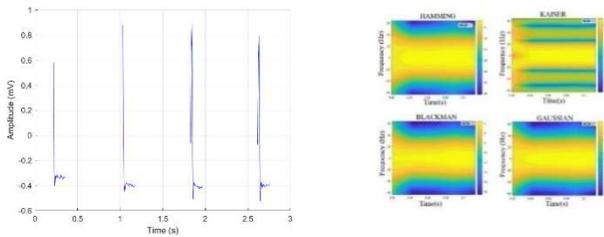

Figure 17. ST Segment of ECG Signal

| Segment | Window Type | Observed Visualization/Patterns | Possible Diseases/Anomalies |
|---|---|---|---|
| PQ | Hamming | Gradual energy increase; stable patterns | Normal atrial activity. |
| | KAISER | Slight energy distortion at edges | Potential minor atrial irregularity. |
| | Blackman | Clear transitions, reduced noise | Normal atrial signal. |
| | Gaussian | Smooth and consistent energy levels | Normal PQ signal; minimal artifacts. |
| QRS | Hamming | High energy at sharp frequencies | Normal ventricular depolarization. |
| | KAISER | Slight loss of clarity in peaks | Possible minor ventricular issues. |
| | Blackman | Well-defined frequency peaks | Normal QRS segment. |
| | Gaussian | Slightly smoothed peak transitions | Normal QRS signal. |
| ST | Hamming | Uniform distribution, stable patterns | Normal ST segment. |
| | KAISER | Smooth but slightly distorted energy | Normal ST segment. |
| | Blackman | Enhanced spectral clarity | Normal ST signal |
| | Gaussian | Minimal spectral leakage, smooth | Normal ST segment; effective filtering. |

| Features | ECG |
|---|---|
| Mean | -0.001320 mv |
| SDSD | 0.164575 mv |
| HBR | 74.29 bpm |

Table 7. Features of ECG Signal